\documentclass{article}

\usepackage{arxiv}

\usepackage[utf8]{inputenc} 
\usepackage[T1]{fontenc}    
\usepackage{hyperref}       
\usepackage{url}            
\usepackage{booktabs}       
\usepackage{amsfonts}       
\usepackage{nicefrac}       
\usepackage{microtype}      
\usepackage{lipsum}
\usepackage{graphicx}
\usepackage{enumitem}
\graphicspath{ {./figures/} }

\title{Scaling human judgment in Community Notes \\with LLMs}

\author{
 Haiwen Li \\
Massachusetts Institute of Technology\\
   \And
 Soham De \\
  University of Washington\\
  \And
 Manon Revel \\
  Berkman Klein Center, Harvard University\\
  \And
 Andreas Haupt \\
  Stanford University\\
  \And
 Brad Miller \\
  X Community Notes\\
  \And
 Keith Coleman \\
  X Community Notes\\
  \And
 Jay Baxter \\
  X Community Notes\\
  \And
 Martin Saveski \\
  University of Washington\\
  \And
 Michiel A. Bakker \\
  Massachusetts Institute of Technology\\
}

\begin{document}
\maketitle
\begin{abstract}
This paper argues for a new paradigm for Community Notes in the LLM era: an open ecosystem where both humans and LLMs can write notes, and the decision of which notes are helpful enough to show remains in the hands of humans. This approach can accelerate the delivery of notes, while maintaining trust and legitimacy through Community Notes’ foundational principle: A community of diverse human raters collectively serve as the ultimate evaluator and arbiter of what is helpful. Further, the feedback from this diverse community can be used to improve LLMs’ ability to produce accurate, unbiased, broadly helpful notes—what we term Reinforcement Learning from Community Feedback (RLCF). This becomes a two-way street: LLMs serve as an asset to humans—helping deliver context quickly and with minimal effort—while human feedback, in turn, enhances the performance of LLMs. This paper describes how such a system can work, its benefits, key new risks and challenges it introduces, and a research agenda to solve those challenges and realize the potential of this approach.
\end{abstract}


\section{Introduction}
Professional fact-checking, which has played a large role in the historical online information environment, is often slow, limited in scale, and lacks trust by large segments of the public \cite{flamini2019most, walker2019republicans, zhao2023insights}. Community Notes (CN) offers a promising counterpart to professional fact-checking. It allows the community to add and rate contextual ``notes’’ on social media posts, shifting the locus of authority from a centralized entity to a decentralized collective. Its strength lies in its ``bridging algorithm’’\footnote{In more detail, users can rate a note as \textit{helpful}, \textit{somewhat helpful}, or \textit{not helpful} and select optional tags to explain their rating. To determine whether to publish a note as \textit{helpful}, The CN algorithm assigns each note a ``helpfulness score’’ based on the ratings it received. The core component of this algorithm is a matrix factorization which predicts each rating as $r_{un} = \mu + i_{u} + i_{n} + f_u \cdot f_n$. The global intercept $\mu$ accounts for the overall propensity of raters to rate notes \textit{helpful}; the user intercept $i_u$ accounts for the baseline propensity of a user to rate notes \textit{helpful}; the note and the user factors $f_u \cdot f_n$ explain the preference of raters with certain viewpoints for certain notes; and the note intercept $i_n$ is treated as the note helpfulness score. This matrix factorization encourages a representation that uses user and note factors to explain as much variation in the ratings as possible, and therefore, a note needs to be universally appealing to achieve a high intercept term/helpfulness score \cite{wojcik2022birdwatch}.}, which surfaces only those notes deemed helpful by raters who typically disagreed in past ratings, ensuring that surfaced context is broadly appealing and thus likely to inform people’s understandings even on contentious topics \cite{wojcik2022birdwatch}. X first launched its ``Community Notes’’ program in 2021 and it has become widespread. \footnote{\href{https://communitynotes.x.com/guide/en/about/introduction}{https://communitynotes.x.com/guide/en/about/introduction}} Meta announced a switch from content moderation with third-party fact-checkers to Community Notes and now makes them available on Facebook, Instagram, and Threads\footnote{\href{https://transparency.meta.com/features/community-notes}{https://transparency.meta.com/features/community-notes}}, YouTube is testing its ``Community Notes’’ feature on videos\footnote{\href{https://blog.youtube/news-and-events/new-ways-to-offer-viewers-more-context/}{https://blog.youtube/news-and-events/new-ways-to-offer-viewers-more-context/}}, and TikTok is rolling out a similar feature under the name ``Footnotes’’.\footnote{\href{https://newsroom.tiktok.com/en-us/footnotes}{https://newsroom.tiktok.com/en-us/footnotes}} Throughout this paper, we use ``Community Notes’’ (CN) to denote this crowd-sourced, community-based fact-checking system, and our discussion applies to all such implementations.

This successful model \cite{slaughter2025community,drolsbach2024community,chuai2024did}, however, is poised for a potential step-change transformation. The core tasks of a Community Notes contributor—researching a claim, synthesizing diverse sources, and drafting a neutral, well-evidenced summary—are capabilities at which ``Deep Research’’ LLMs show promise \cite{geminiDeepResearch,openaiDeepResearch,xaigrok}. Indeed, work from 2024 has demonstrated that a fully automated pipeline for generating Community Notes with LLMs can, under certain circumstances, produce notes that are of similar quality to human-written notes—at a fraction of the time and effort \cite{zhou2024correcting}. LLM-written CNs have the potential to be faster to produce, less effort to generate, and of high quality, hence are an attractive direction to pursue.

While LLM-written CNs are compelling, critical questions remain: Can LLMs consistently produce accurate notes that are well-received across perspectives? If yes, can we use them to accelerate the addition of informative context in a way that is valued and trusted across viewpoints, while avoiding the fate of becoming just a new version of top-down fact-checking that can lack broad trust? We believe both are possible.

Rather than replacing humans, LLMs can complement and enhance their work. Trust in CN stems not from who drafts the notes, but from who evaluates them. It is the collective judgment of a diverse and engaged community that grounds people’s trust in CN. 

This paper argues for a new paradigm for Community Notes in the LLM era: an open ecosystem where notes from both human writers and LLMs are submitted into a single pool, and the decision of which notes are helpful enough to show remains in the hands of the people. The system’s legitimacy is still upheld by its foundational principle: a community of diverse human raters that collectively serve as the ultimate evaluator and arbiter of what is helpful. Further, the feedback from this diverse community can be used to improve LLMs’ ability to produce accurate and broadly helpful notes. In this way, LLMs are both an asset to humans—helping deliver context quickly and with less effort—and those humans’ feedback is an asset that further improves the work those LLMs are doing. A virtuous cycle.

Integrating LLMs into the CN framework represents a transformative opportunity, one that offers promise of dramatically increased scale and speed of its contextual notes. However, this integration is not a simple technical change; it introduces a new set of high-stakes challenges. This paper aims to provide a guide for navigating this complex landscape. We will outline the benefits of this hybrid human-AI model, critically examine the associated risks, and propose a concrete research agenda designed to harness the power of LLMs responsibly while reinforcing the core principles of community-led approach.

\section{Proposed hybrid system: Human and LLM note writing with human-only rating}
We propose an expansion of the Community Notes pipeline, establishing a clear and synergistic division of labor between AI and the human community. 

In this new model, both humans and LLMs can contribute as note writers, while only humans rate notes and ultimately determine which notes are helpful enough to show. It seems likely that the combination of human and LLM writers is superior to either on its own, both in terms of generating trust and producing more effective notes. Humans and LLMs can bring different and additive strengths to the writing process.

\begin{figure}[ht!]
  \centering
  \includegraphics[width=\textwidth]{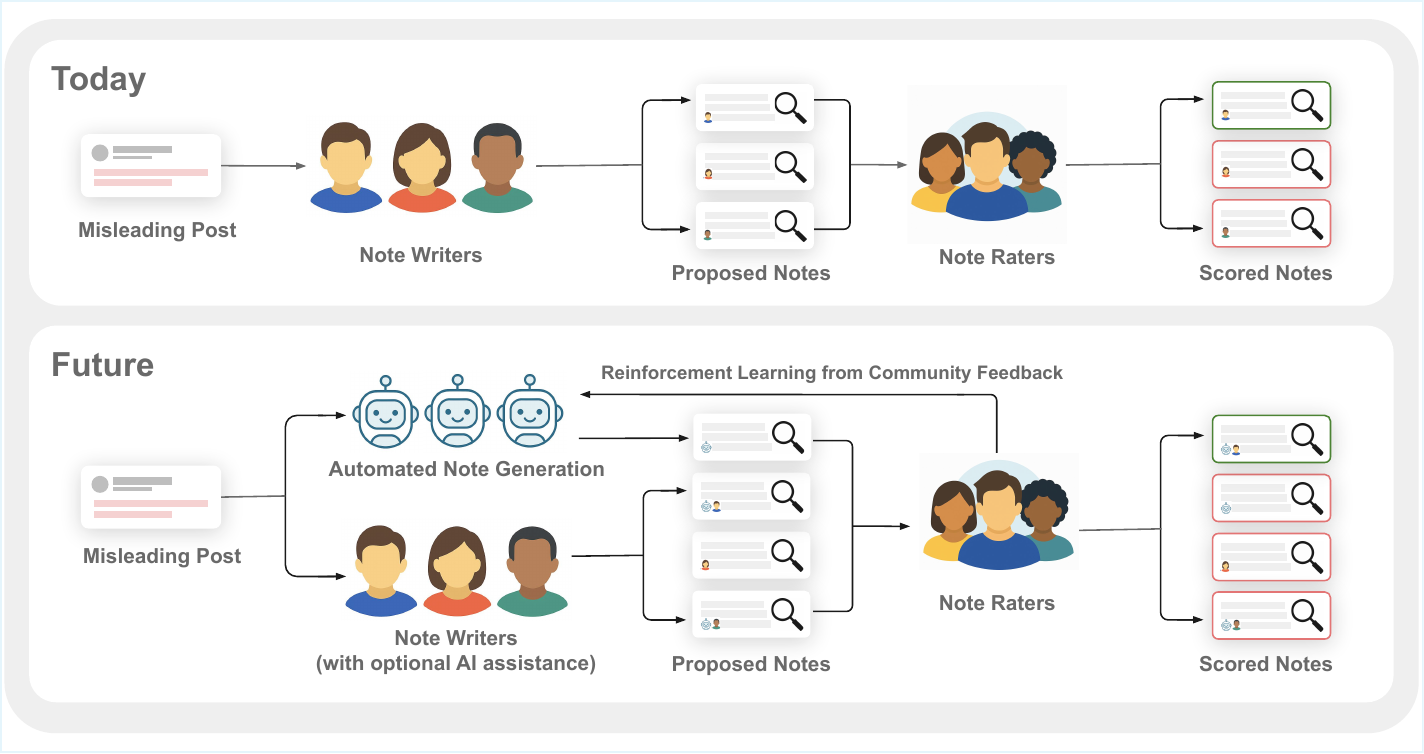}
  \caption{\textbf{An expansion of the Community Notes pipeline from ``all-human’’ to a hybrid ``human-LLM’’ model.} Today (top): Human note writers draft proposed notes in response to a misleading post, and other human contributors rate their helpfulness; the bridging algorithm picks the broadly helpful notes. Future (bottom): LLMs will also participate in the writing stage, producing notes or assisting human writers, while the rating stage remains human-only. Community feedback from human raters flows back to improve LLM note generation (RLCF).
 }
  \label{fig:future}
\end{figure}

\newpage
\subsection{Notes may be proposed by either humans or LLMs}
\subsubsection{Automated note creation by LLMs}
Allowing automated note creation would enable the system to operate at a scale and speed that is impossible for human writers, potentially providing context for orders of magnitude more content across the web. There are multiple stages which can be automated:
\begin{itemize}
    \item \textbf{Post Selection:} Identifying which content warrants a note, which can involve prioritizing:
    \begin{enumerate}[label=\alph*.]
        \item \textbf{Content with high potential reach:} Posts algorithmically predicted to gain significant viewership or virality, maximizing the note's impact.
        \item \textbf{Content likely to be misleading:} Posts where a note has a strong potential to be marked helpful by the bridging algorithm.
        \item \textbf{Community-flagged content:} Posts flagged by human users as needing context, potentially accompanied by initial evidence, first-hand observations, or specific insights provided by the flagger.
    \end{enumerate}
    \item \textbf{Research:} Often including search and tool-use to gather information from a wide array of verifiable sources.
    \item \textbf{Evidence synthesis:} Analyzing and synthesizing conflicting or complementary pieces of information to form a coherent factual basis.
    \item \textbf{Note composition:} Drafting a clear, neutral, and well-sourced note designed to provide helpful context.
\end{itemize}
While the primary benefits of automated note creation are likely to be speed and scale, there may be other benefits too: for example, LLMs can learn directly from the collective ratings of the community to internalize what makes a note helpful. As we will discuss in detail later, this process—which we term Reinforcement Learning from Community Feedback (RLCF)—may allow for more consistent adaptation than is possible for human writers. 

\subsubsection{Human writers}
In this hybrid system, human contributors could write notes from scratch or utilize AI assistance to help draft and source their contributions. Humans can benefit from AI assistance but not have to defer to it — they can write notes where they want (for example where LLMs aren’t aware of the need) or improve upon sub-par LLM notes.

Human CN writers have demonstrated proficiency at handling novel (e.g. breaking news) situations where available information is limited and changing rapidly, niche topics where they are able to contribute deep knowledge on specific subjects, or scenarios where it’s essential to understand how a post might be interpreted by people in order to understand why or why not a note might be valued.

Further, they are uniquely positioned to identify deficits or biases in the LLMs by writing notes on topics the automated writers overlook. They can help explore novel applications for notes: for example, although Community Notes was originally intended primarily as a way to add context to potentially misleading posts, some contributors started using it to do other types of crowd-sourced moderation such as flagging spam, scams, or crediting original creators of copy-pasted posts.

\subsection{Unified evaluation: Human raters as the sole arbiters}
While we advocate for contributions of LLMs for note authorship, we believe that evaluation from users is crucial. All notes, whether written by a human or an LLM, enter the same rating pool. The human community remains the ultimate arbiter of what gets published. Contributors' ratings are the primary input into the core approval algorithm. Currently, raters may also attach optional tags (e.g., ``Incorrect information’’, ``Argumentative or biased language’’) along with their ratings; to make this process more dynamic, the platform could allow additional free-text explanations from raters (e.g., ``The note resonates with my experience, though the cited statistic feels outdated’’). These richer critiques can guide LLMs to generate a revised version immediately, and serve as valuable training signals for improving future note generation.   

The bridging algorithm will continue to be the central mechanism for ensuring that only ``bridging’’ notes—those found helpful by raters with diverse perspectives—are surfaced. This structure ensures that the system's trust and legitimacy do not derive from the writer (human or LLM) but from the collective curation and pluralistic judgment of the human community.

\section{Benefits of this hybrid system}
This proposed human-AI partnership offers profound benefits over a purely human (or purely AI) system:

\begin{itemize}
    \item \textbf{Scalability and speed:} The automated LLM writing pipeline can respond to misleading content in real-time and address a far greater volume of content than a human-powered system, tackling the ``long tail’’ of niche misleading claims and out-of-context stories.
    \item \textbf{Enhanced quality:} The hybrid model fosters a higher standard of quality through several mechanisms. First, by having human- and LLM-written notes compete in the same rating system, a natural selection process emerges where the most helpful context—as determined by the community—rises to the top. Second, the system benefits from the complementary strengths of both humans and AI. LLMs excel at rapidly synthesizing widely available information, while human contributors are uniquely skilled at addressing novel situations and providing niche expertise. Finally, this partnership can enhance the pluralism of options presented to the community. If LLMs generate initial drafts that represent a wider range of perspectives than a single human writer typically could, the quality of community deliberation is improved from the start.
    \item \textbf{Preserve participatory oversight:} Human oversight is maintained because human collective judgment is elicited universally on all notes. The system’s legitimacy is upheld because the diverse human community remains the ultimate arbiter, collectively deciding what context is helpful enough to be shown, regardless of its origin.
\end{itemize}
\newpage
\section{Risks and challenges}
This new approach introduces a new set of high-stakes challenges that must be proactively addressed.
\begin{itemize}
    \item \textbf{Persuasive but inaccurate notes:} LLMs optimized to maximize helpfulness scores may become exceptionally skilled at crafting persuasive, emotionally resonant, and seemingly neutral notes that might not be factually accurate, lead to ``helpfulness hacking’’ \cite{sharma2023towards, bondarenko2025demonstrating, schoenegger2025large}. If rated helpfulness isn't perfectly correlated with accuracy, then highly-polished but misleading notes could be more likely to pass the approval threshold. This risk could grow as LLMs advance; they could not only write persuasively but also more easily research and construct a seemingly robust body of evidence for nearly any claim, regardless of its veracity, making it even harder for human raters to spot deception or errors.
    \item \textbf{Impact on human writer engagement:} Even with the ability to write, will human contributors remain motivated if LLMs are generating well-researched notes at scale? It is conceivable that if the platform becomes dominated by the speed and scale of AI, the sense of ownership and ``skin in the game’’ that drives community participation could diminish \cite{burtch2024consequences, wagner2025death}, threatening the pool of diverse writers (and perhaps indirectly also raters) essential for the system's long-term health and legitimacy.
    \item \textbf{Homogenization and reduced creativity:} While human writers provide some safeguard against total uniformity, a risk remains that the LLM-generated portion of the ecosystem will converge on a single, bland, ``optimally inoffensive’’ style to pass the bridging algorithm. This could not only create a sterile information environment but also subtly devalue the more expressive or nuanced styles of human writers, crowding out the very creativity the hybrid model is meant to preserve.
    \item \textbf{Risk of rater overload:} In a world with a very large volume of notes, how do we ensure human rating capacity can sufficiently scale? LLMs could strategically select where to write notes (prioritizing high-reach, likely misleading, or community-flagged content) in order to maximize impact, yet even so the sheer capacity of LLMs to generate notes for all such identified content could lead to a significant increase in the total volume of notes entering the rating system. This higher volume of notes could overwhelm the capacity of human raters or dilute the impact of the most genuinely critical notes if raters are spread too thin across a larger pool of notes.
\end{itemize}

\section{A Research agenda for LLM-Powered Community Notes}
To successfully build a thriving ecosystem, we propose a focused research agenda. 
\begin{figure}[h!]
  \centering
  \includegraphics[width=\textwidth]{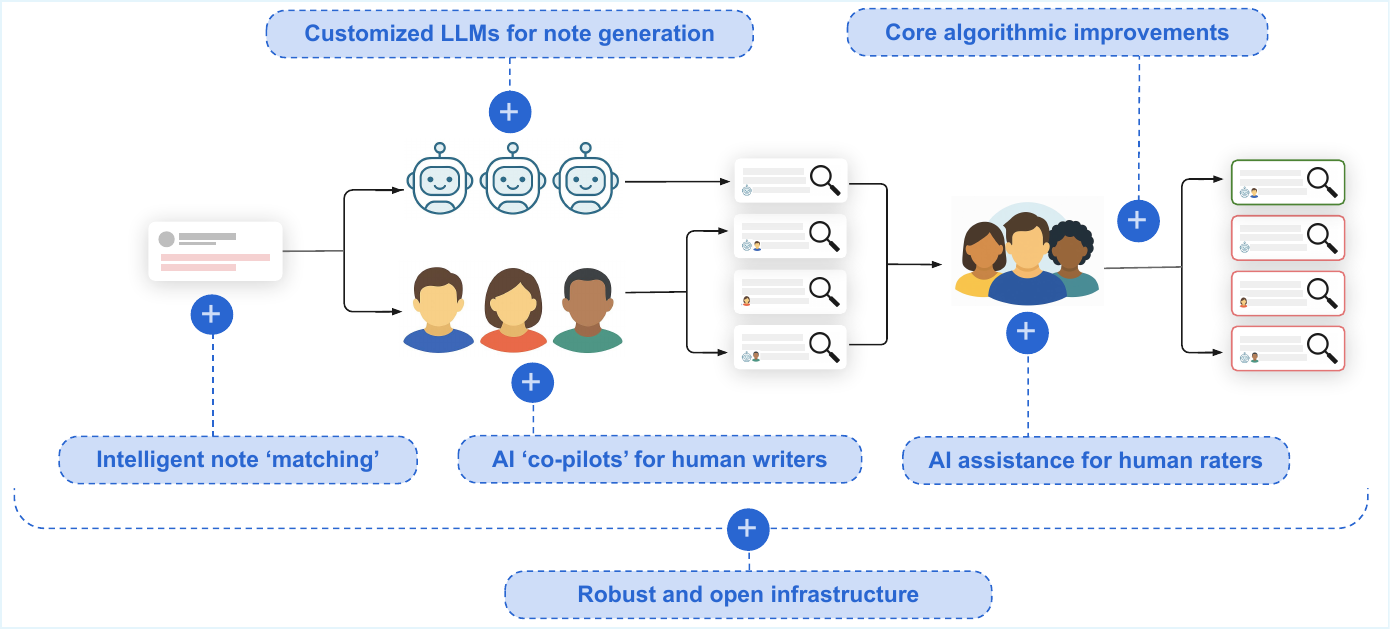}
  \caption{\textbf{A research agenda for LLM-powered Community Notes.} (1) Customized LLMs for note generation;  (2) AI `co-pilots’ for human writers; (3) AI assistance for human raters; (4) Intelligent note `matching’ adapts existing helpful notes to new, similar contexts; (5) Evolving the core algorithm for AI-generated content; (6) Building a robust and open infrastructure. }
  \label{fig:research}
\end{figure}

\subsection*{1. Customized LLMs for note generation:}
The rich, informed judgments gathered through human oversight provide the ideal foundation for a constructive cycle of improvement. While sophisticated prompting of ``Deep Research’’ models show promise at producing high-quality initial notes, a significant opportunity lies in developing specialized post-training methods to further optimize them for the unique demands of the Community Notes ecosystem. One promising direction could be to adapt the paradigm of Reinforcement Learning from Human Feedback (RLHF) to this unique community context, a process we can term Reinforcement Learning from Community Feedback (RLCF).

Unlike traditional RLHF \cite{ziegler2019fine, ouyang2022training}, which often relies on simple pairwise preferences from a small group of labelers, RLCF would leverage the pluralistic signal of the entire community. The technical implementation would be more sophisticated: instead of predicting a final ``bridging’’ score directly, a reward model could be trained to predict the likely ratings of various diverse types of users, similar to \cite{de2025supernotes}. From these simulated ratings, the system could then aggregate them to calculate a note’s expected intercept score ($i_n$), which represents its predicted ability to bridge divides. This intercept would then serve as the core reward signal to fine-tune the LLM.

However, a naive implementation that only maximizes this predicted score would lead directly to the ``homogenization’’ risk, as the model would quickly find a safe, bland optimum. This opens a critical research frontier: designing RLCF frameworks that balance two crucial objectives. The system must not only refine strategies already known to work, but also actively encourage experimentation to discover novel, potentially superior, approaches. For instance, the system could reward LLMs not just for high predicted scores, but also for generating notes that are stylistically novel or that explore different dimensions in their framing or research, learning from the human-authored examples.

The ultimate goal is to use the community's feedback not just to ratify known good notes, but to discover new, potentially even better, ways of creating context. More generally, RLCF represents just one path forward. The open and rich dataset of notes, ratings, and feedback from Community Notes creates a unique testbed for developing entirely new post-training methods. Future research could explore models that are explicitly trained to leverage tool use, such as calling external data APIs or performing automated source verification, as part of their generation process, learning directly from this open data. This opens the door to creating a new class of transparent, verifiable, and pluralistically aligned models tailored for public discourse \cite{sorensen2024roadmap}.

\subsection*{2. AI co-pilots for human writers:}
In the hybrid model, human writers are an essential part of the ecosystem. One potentially valuable research direction, therefore, is to build powerful AI-assisted writing tools that act as co-pilots. These tools could help human writers research topics faster, suggest relevant sources, generate initial drafts for editing, or polish the tone and clarity of a final note. This directly supports the human role as an innovator while also introducing an important challenge: how to empower human writers with co-pilots without flattening style and voice into homogeneous, bland writing. Studying this co-writing process will also provide invaluable data on how to best support and grow human contribution.

\subsection*{3. AI assistance for human raters:}
AI-powered tools could help human raters audit note claims more efficiently, and prevent them from rating a note solely based on its surface-level persuasiveness \cite{saunders2022self, bridgers2024}. While there is a potential risk that these tools could be erroneous in the same way LLM note writers may be, there exist clever strategies to help mitigate it. For example, one concept would be to task multiple AIs to adversarially debate the merits of a note \cite{irving2018ai, michael2023debate}: one AI focuses on critiquing the note, checking its sources for misrepresentation, searching for counter-evidence, and challenging the logical consistency of its argument, while another AI defends it. This adversarial process could help instantly surface potential flaws, hidden biases, or fabricated evidence, empowering the human rater to make a more informed judgment.  Instead of starting from scratch, the rater now plays the role of an adjudicator—evaluating a structured clash of arguments. In the AI safety literature, this approach is known as scalable oversight: a strategy for amplifying human judgment through structured assistance, enabling reliable supervision of systems that may surpass human capabilities in speed or rhetorical fluency \cite{irving2018ai,kenton2024scalable}.

\subsection*{4. Intelligent note matching and adaptation for new contexts:}
A critical bottleneck in the hybrid system is the potential for AI-generated notes to overwhelm the finite rating capacity of the human community. A promising research direction to mitigate this ``rating-demand’’ imbalance is to develop methods for intelligently adapting and reapplying notes that have already been validated by human raters to new, similar contexts. Given that a claim may reappear verbatim, or with slight rephrasing, or attached to different media, this is a non-trivial challenge. Automatically matching notes to posts that people do not think need them could significantly undermine trust in the system. Research is needed to explore how to best leverage existing human-approved judgments. Key questions include:
\begin{itemize}
    \item \textbf{Contextual matching:} How can we develop robust models to determine if a new post is semantically similar enough to an existing post (in its claim, media, or context) to warrant reusing an existing note, keeping in mind that sufficient similarity is a matter of rater perception in context?
    \item \textbf{AI-powered adaptation:} What are the safe boundaries for AI-driven modification? Can an AI autonomously make minor edits to a note to better fit a new context, or should it only propose changes for human review?
    \item \textbf{User experience and governance:} How should these adapted notes be presented to users? Should they be explicitly labeled as ``adapted from a previously rated note’’? What new rating mechanisms are needed? For example, should users be able to downvote the match itself, separate from the note's content, if they feel the adaptation is inappropriate?
\end{itemize}
Successfully developing such a system would dramatically amplify the impact of every human rating, allowing the community's judgment to scale more effectively across the information ecosystem.

\subsection*{5. Evolving the core algorithm for AI-generated content:}
The current ``bridging’’ algorithm is calibrated on the idiosyncrasies of human-written notes. A flood of highly optimized, grammatically perfect LLM-generated notes could change the rating dynamics. Research will help determine to what degree the distribution of content shifts with LLM note writers, and whether it could benefit from algorithmic adaptation. This might involve incorporating metadata about a note’s origin (human, AI-assisted, or fully AI) or adjusting model parameters to prevent the stylistic polish of LLM notes from being overweighted against the substantive, nuanced contributions of human writers.

\subsection*{6. Building a robust and open infrastructure:}
A healthy ecosystem requires a new kind of infrastructure.
\begin{itemize}
    \item \textbf{Human rater verification:} Integrity of the entire system is strengthened by the authenticity of its human participants and their engagements. The system should robustly work to increase the likelihood that ratings are from unique humans, where rating outcomes reflect the genuine viewpoints of people from different perspectives. This could include everything from use of privacy-preserving methods to verify that each rating account is associated with one, distinct human being, to implementing limits to the influence any single account can have, to detecting and limiting the impact of accounts attempting to coordinate to influence outcomes. Such approaches have already been deployed on Community Notes systems, and can continue to evolve and expand \footnote{\href{https://communitynotes.x.com/guide/en/contributing/signing-up}{https://communitynotes.x.com/guide/en/contributing/signing-up}}\textsuperscript{,}\footnote{\href{https://about.fb.com/news/2025/03/testing-begins-community-notes-facebook-instagram-threads/}{https://about.fb.com/news/2025/03/testing-begins-community-notes-facebook-instagram-threads/}}.
    \item \textbf{An open note-writing API:} To foster innovation, trust, and maintain the open, participatory nature of Community Notes, platforms could create an open API that allows any developer or organization to submit their LLM-generated notes for a given post. These notes would all enter the same rating pool, competing on a level playing field. The community's ratings would act as a natural selection mechanism, rewarding models that produce the most genuinely helpful content.
    \item \textbf{Prescreening and diversity curation:} As open APIs enable a higher volume of note submissions, one potential optimization is to introduce a prescreening layer that helps manage rater attention. This system would operate at two levels. First, it needs to strategically allocate rater attention across posts. As AI makes it possible for many more posts to have note candidates, the system must be more selective in ``nudging’’ them to engage with the most impactful or contentious notes and posts. Second, for any given post, this layer could utilize techniques like semantic clustering to group nearly identical notes  and select a small, diverse set for evaluation, ensuring raters are presented with meaningfully different options rather than five variations of the same argument. By managing attention at both the post-level and the note-level, this system could directly mitigate the risk of rater overload.
\end{itemize}

\section{Long-term vision and conclusion}
The hybrid model we propose is not just an upgrade to Community Notes, but a blueprint for a new form of human-AI collaboration in the production of public knowledge. This extends beyond simple fact-checking into a dynamic, positive feedback loop. Imagine a system where humans contribute unique, first-hand observations that LLMs cannot access—photos from a local event, specialized domain knowledge, or on-the-ground reports. LLMs are then tasked with verifying this information against public sources and integrating it into a comprehensive, well-structured note. The human community, in turn, evaluates not just the note's helpfulness but the fidelity of the LLM integration, creating a robust, self-correcting engine for generating trusted public information.

This knowledge-production engine can evolve beyond a single platform into a dynamic, contextualization layer for the entire web. Such a layer could operate on specific URLs or, instead, it could operate on the content itself, using techniques like fuzzy matching or semantic hashing to attach context to claims, images, or videos as they spread across different domains. Imagine a browser-level system that analyzes the content of any page you visit and, drawing from this universal, continuously rated stream of notes, provides a non-intrusive overlay with relevant background, fact-checks, or warnings about a source's bias.

Its ultimate importance becomes clear when we consider the trajectory of the internet itself. We are heading toward a future of radical personalization, where the very concept of a common information space may dissolve \cite{flaxman2016filter, rodilosso2024filter, nehring2024large}. Imagine a world where our primary interface to information is not a browser fetching distinct web pages, but a personal AI agent that synthesizes and generates content just for us—a personalized news feed, a summarized search result, a conversational answer. Every piece of content is mediated and optimized to align with our individual preferences, biases, and existing beliefs.

In such a future, where every individual's information diet is perfectly tailored, the role of a universal, non-personalized system like Community Notes could become even more important. Rather than simply being a feature on a single platform, CN could evolve into part of the shared infrastructure that helps maintain a common information space. While personal AIs may optimize content for individual preferences, Community Notes works in the opposite direction: its bridging algorithm is designed to surface only the context that people from different perspectives agree is helpful.

In that world, CN could serve as a public, transparent, and pluralistic anchor—complementing private, personalized feeds with content that reflects a broader consensus. It wouldn’t enforce a singular truth, but it could help illuminate the contours of agreement. This makes CN not just a tool for contextualizing posts, but a potential foundation for pluralistic AI alignment—one built on human deliberation, not just individual optimization.

The goal is not to create an AI assistant that tells users what to think, but to build an ecosystem that empowers humans to think more critically and understand the world better. At the heart of this paper is a simple idea: LLMs and humans can work together in a virtuous loop. LLMs help surface context faster, cheaper, and at greater scale while humans rate which notes are actually helpful. That pluralistic feedback can then be used to make LLMs better at producing the kind of context people value most. This feedback loop—Reinforcement Learning from Community Feedback (RLCF)—has the potential to improve both sides: giving people more trusted information, and making LLMs fundamentally better at providing it.

\section*{Acknowledgements}
We are grateful to Stephanie Chan, Jiansong Chao, Sophie Hilgard, Rishub Jain, and Daniel Ortiz for helpful conversations and feedback on earlier drafts.

\bibliographystyle{unsrt}  
\bibliography{references}  






\end{document}